\documentclass{aastex}          
\usepackage{spr-astr-addons}     
\usepackage{url}

\RequirePackage{color}

\usepackage[breaklinks=true]{hyperref}
\hypersetup{colorlinks,linkcolor=blue,citecolor=blue,urlcolor=blue}
\usepackage{txfonts}

\begin{document}
%
\title{Structure of IRAS\,05168+3634 star-forming region}

\shorttitle{Structure of IRAS 05168+3634}
\shortauthors{Nikoghosyan et al.}

\author{Nikoghosyan E. H. \altaffilmark{}} \and \author {Azatyan N. M.\altaffilmark{}} \and \author {Andreasyan D. H.\altaffilmark{}} \and \author {Baghdasaryan D. S.\altaffilmark{}}
\affil{Byurakan Astrophysical Observatory, 0213, Aragatsotn prov., Armenia}
\email{elena@bao.sci.am} 


\begin{abstract}
This study aims to determine the main physical parameters (N(H$_2$) hydrogen column density and T$_d$ dust temperature) of the Interstellar medium, and their distribution in the extended star-forming region, which includes IRAS\,05156+3643, 05162+3639, 05168+3634, 05177+3636, and 05184+3635 sources. We also provide a comparative analysis of the properties of the Interstellar medium and young stellar objects. Analysis of the results revealed that Interstellar medium forms relativity dense condensations around IRAS sources, which are interconnected by a filament structure. In general, in sub-regions T$_d$ varies from 11 to 24\,K, and N(H$_2$) - from 1.0 to 4.0\,$\times$\,10$^{23}$\,cm$^{-2}$. The masses of the ISM vary from 1.7\,$\times$\,10$^4$ to 2.1\,$\times$\,10$^5$ M$_{\odot}$. All BGPSv2 objects identified in this star-forming region are located at the N(H$_2$) maximum. The direction of the outflows, which were found in two sub-regions, IRAS\,05168+3634 and 05184+3635, correlates well with the isodenses' direction. The sub-regions with the highest N(H$_2$) and Interstellar medium mass have the largest percentage of young stellar objects with Class\,I evolutionary stage. The wide spread of the evolutionary ages of stars in all sub-regions (from 10$^5$ to 10$^7$ years) suggests that the process of star formation in the considered region is sequential. In those sub-regions where the mass of the initial, parent molecular cloud is larger, this process is likely to proceed more actively. On the \textit{Gaia}\,EDR3 database, it can be assumed that all sub-regions are embedded in the single molecular cloud and belong to the same star-forming region, which is located at a distance of $\sim$\,1.9\,kpc.
\end{abstract}



\section{Introduction}
\label{s:1}
It is known that star formation is an intrinsically complex process involving the collapse and accretion of matter onto protostellar objects, but also the mass loss from the star-forming system in the form of bipolar outflows or stellar wind \citep{Lada2003}. Therefore, there is a genetic connection between young stellar objects (YSOs) and surrounding gas-dust matter \citep[e.g.][]{Gonz2020}. This necessitates an integrated approach to study star-forming regions. The integrated approach implies a detailed study and determination of the main properties of already formed young stellar clusters (density, mass function, evolutionary age distribution, etc.) and the environment (density, temperature, chemical composition, maser emission, etc.). Moreover, embedded stellar clusters can many tell us about the initial star formation scenarios. For example, if the star formation is triggered, the age spread of new generation stars should be small, while in self-initiated condensations the age spread of young stellar clusters is large \citep[e.g.][]{Zinnecker1993,Preibisch2012}. Consequently, an integrated approach is needed to understand the holistic picture of the star formation process, which includes the study of both ISM and YSOs.

This work aims to study the Interstellar medium (ISM) of the molecular cloud and search for evidence of a relation between the properties of YSOs and their parent molecular cloud in the extended star-forming region associated with IRAS\,05168+3634 source, which is also known as Mol 9 \citep{Molinari1996}. There are different manifestations of star formation activity in the region, associated with this IRAS source: H$_2$O, NH$_3$, 44\,GHz CH$_3$OH, and OH maser emission \citep{Zhang2005,Molinari1996,Ruiz-Velasco2016,Fontani2010}, CS emission \citep{Bronfman1996}, SiO (J\,=\,2 $-$ 1) line emission \citep{Harju1998}, $^{13}$CO cores \citep{Guan2008}, etc. \citet{Zhang2005} have discovered a molecular outflow in this region, and \citet{Wolf-Chase2017} have detected collimated outflows of near-infrared Molecular Hydrogen emission-line Objects (MHOs).

According to \citet{Sakai2012} the trigonometric parallax of IRAS\,05168+3634 is 0.532\,$\pm$\,0.053\,mas, which corresponds to a distance of 1.88$^{+0.21}_{-0.17}$ kpc. However, a kinematic distance was estimated at 6.08\,kpc \citep{Molinari1996}. Taking into account the high accuracy of direct parallax measurement, in our work all calculations were performed for this distance estimate.

The embedded stellar cluster, associated with IRAS 05168+3634 star-forming region has been investigated by various authors in the near-infrared (NIR) and middle-infrared (MIR) wavelengths \citep{Kumar200,Faustini2009}. \citet{Wang2009} suggested that in the vicinity of IRAS\,05168+3634 source there is a high-mass star-forming region in the pre-UC\,HII phase. \citet{Azatyan2016} have shown that in the star-forming region a bimodal cluster is located. Moreover, a more extensive and detailed study shown that IRAS\,05168+3634 source is embedded in the extended molecular cloud, which includes four additional star-forming sub-regions, associated with IRAS\,05156+3643, 05162+3639, 05177+3636, and 05184+3635 sources \citep[][hereafter A19]{Azatyan2019}. Totally, in the molecular cloud, it was identified 240 YSOs, with evolutionary ages from 0.1 to 3\,Myr, and masses from 0.2 to 7\,M$_{\odot}$ (A19). It should be noted that, excluding IRAS\,05168+3634, there is very little information about other IRAS sources and their surrounding ISM. 

The paper presents the results of a study of ISM properties, including hydrogen column density and dust temperature in IRAS 05168+3634 extended star-forming region. We also provide a comparative analysis of the properties of the ISM and YSOs. We have organized the paper in the following way: Section \ref{s:2} describes used data; in Section \ref{ss:3.1} we discuss the properties of the gas–dust matter in the region; in Section \ref{ss:3.2} it is analyzed the properties of each sub-region individually. The obtained results are discussed and summarized in Sections \ref{s:4} and \ref{s:5}, respectively.

\section{Used data}
\label{s:2}
For study gas-dust matter, we used far-infrared (FIR) wavelengths images in the range of 160–500\,$\mu$m, obtained by using the Photodetector Array Camera and Spectrometer \citep{Poglitsch2010}, as well as the Spectral and Photometric Imaging Receiver \citep{Griffin2010} at the 3.5\,m \textit{Herschel} Space Observatory. We used the 2.5 Level PACS and SPIRE maps available in the \textit{Herschel} Science Archive. The observations for PACS (160\,$\mu$m band) and SPIRE (250$-$500\,$\mu$m bands) maps were carried out in parallel mode with fast scanning speed (60\,arcsec/sec). The FWHM beam sizes of the original \textit{Herschel} maps are approximately 12.3, 17.6, 23.9, and 35.2\,arcsec/pixel at 160, 250, 350, and 500\,$\mu$m, respectively.

We also used the images obtained by the Wide-field Infrared Survey Explorer \citep[WISE,][]{Wright2010} in four MIR bands (3.4, 4.6, 12, and 22\,$\mu$m). The angular resolutions of WISE maps are respectively 6.1, 6.4, 6.5, and 12.0\,arcsec/pixel at corresponding WISE bands and have an astrometric precision better than 0.15 sources detected with good S/N. The processed WISE images were acquired via the NASA/IPAC Infrared Science Archive (IRSA).

For checking the distance and proper motion of several objects in the star-forming region, we used the archival data of \textit{Gaia} Data Release 3 \citep[EDR3, \url{https://gea.esac.esa.int/archive/},][]{Gaia2021}, including coordinates, parallaxes, and proper motions. \textit{Gaia} EDR3 contains data of astrometry and photometry for 1.8 billion sources brighter than magnitude 21, complemented with the list of radial velocities from \textit{Gaia} DR2. \textit{Gaia} EDR3 represents a significant advance over \textit{Gaia} DR2, with parallax precisions increased by 30\,\%, proper motion precisions increased by a factor of 2, and the systematic errors in the astrometry suppressed by 30\,$-$\,40\,\%  for the parallaxes and by a factor $\sim$\,2.5 for the proper motions.

\section{Results}
\label{s:3}
\subsection{Dust emission}
\label{ss:3.1}
\begin{figure*} [h]
\centering
\includegraphics[width=1.5\columnwidth]{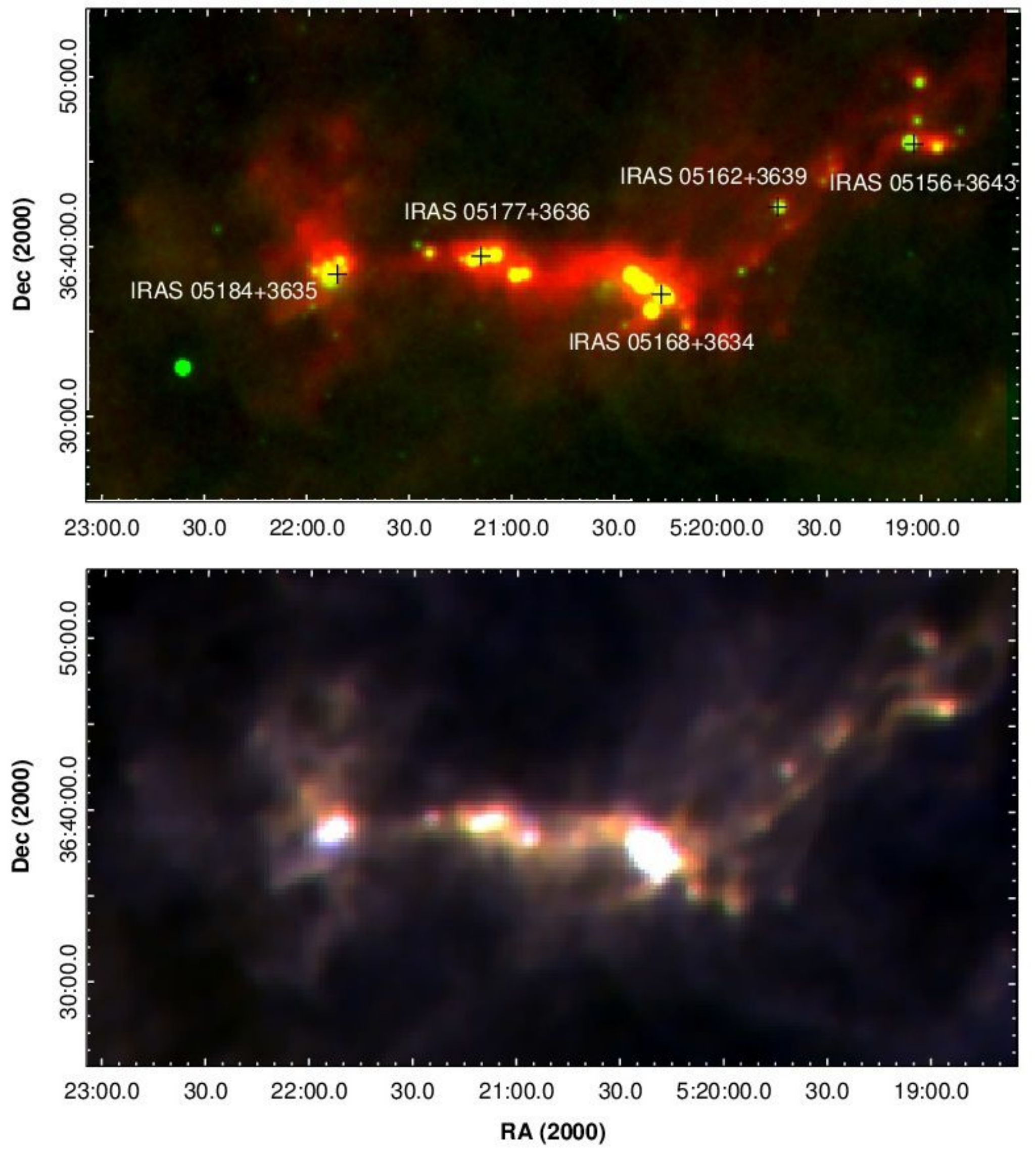}
\caption{Colour-composite images of IRAS\,05168+3634 star-forming region. \textit{Top panel}: W4 22\,$\mu$m (green) and \textit{Herschel} 500\,$\mu$m (red), \textit{bottom panel}: \textit{Herschel} 160\,$\mu$m (blue), 350\,$\mu$m (green), and 500\,$\mu$m (red). The positions of IRAS sources are marked by black crosses.} 
\label{fig:1}
\end{figure*}

As already noted in the Introduction, A19 found that IRAS\,05168+3634 source belongs to an extended star-forming region, which includes 4 more IRAS sources. Around each of them, a group of YSOs with intermediate- and low-mass is formed. The star-forming region is embedded in a molecular cloud that is reflected in the top panel of Figure \ref{fig:1}, where a colour-composite image composed through the WISE 22\,$\mu$m and \textit{Herschel} 500\,$\mu$m bands is presented. Figure \ref{fig:1} shows that stellar objects are well distinguishable in the MIR range, while the continuum emission of the ISM in this range is weak but intense in the FIR. This indicates a relatively low temperature of the ISM. The bottom panel of Figure \ref{fig:1} shows the colour-composite image composed through the \textit{Herschel} 160, 350 and 500\,$\mu$m bands. The gas-dust matter has an inhomogeneous structure, forming well-defined condensations around IRAS sources. Condensations are most pronounced in the vicinity of IRAS 05168+3634, 05177+3636, and 05184+3635. It should be noted that the intensity of radiation of the ISM in all three ranges is almost the same. The condensations are connected by a filament structure, which is better expressed in the longer wavelength range, which indicates its lower temperature.

To obtain the hydrogen column density and dust temperature, the thermal emission from cold dust lying in the {\itshape Herschel} FIR optically thin bands (160–500\,$\mu$m) can be used \citep{Hildebrand1983,Battersby2011}. Following the discussion in the previous studies \citep[e.g.][]{Battersby2011} this wavelength range well applicable for those cases where the dust temperature is in the range of 5-50\,K, while there are no clear restrictions for the column density. Because we excluded the emission of 70\,$\mu$m band, which would have a significant contribution from the warm dust component, and, therefore, modelling with a single-temperature blackbody would over-estimate the derived temperatures. Besides, in this range the optically thin assumption would not hold. 
 
For our task, Level\,2.5 processed {\itshape Herschel} images were downloaded through the {\itshape Herschel} Interactive Processing Environment \citep[HIPE;][]{Ott2010}. To eliminate the effect of bad pixels on the final result, we gave them the median value of eight immediate-neighbour pixels. For the estimation of hydrogen column density and dust temperature, using the convolution kernels of \citet{Aniano2011}, we convolved the 160–350\,$\mu$m images to the resolution of the 500\,$\mu$m image (the lowest among all),  and regridded to a pixel scale of 14\,arcsec (same as at 500\,$\mu$m).  This procedure were carried out using the HIPE software. 

Modified single-temperature blackbody fitting was subsequently carried out on a pixel-by-pixel basis using the following formula:

\begin{equation}
\label{equ:1}
S_\nu = B_\nu(\nu,T_d)\Omega(1-e^{-\tau(\nu)}), \\   
\end{equation}
with
\begin{equation}
\label{equ:2}
\tau(\nu)=\mu_{H_2}m_{H}k_{\nu}N(H_2),
\end{equation}
where $\nu$ is the frequency, $S_\nu(\nu$) is the observed flux density, $B_\nu(\nu, T_d)$ is the Planck function, $T_d$ is the dust temperature, $\Omega$ is the solid angle in steradians from where the flux is obtained (in this case for all bands the solid angle subtended by a 14\,arcsec\,$\times$\,14\,arcsec pixel), $\tau(\nu)$ is the optical depth, $\mu_{H_2}$ is the mean molecular weight (adopted as 2.8 here), $m_H$ is the mass of hydrogen, $k_\nu$ is the dust opacity, and $N(H_2)$ is the hydrogen column density. For opacity, we adopted a functional form of $k_{\nu}= 0.1(\nu/1000\, GHz)^{\beta}\,cm^2 g^{-1}$, with $\beta=2$ \citep[see][]{Hildebrand1983,Andr2010}. For each pixel, equation (\ref{equ:1}) was fitted using the four data points (160, 250, 350, and 500\,$\mu$m) keeping T$_d$ and N(H$_2$) as free parameters. \citet{Launhardt2013} used a conservative 15$\%$ uncertainty in the flux densities of the {\itshape Herschel} bands. We adopted the same value here for all bands. The uncertainties of the parameters were derived using Pearson’s $\chi^2$ statistics:
\begin{equation}
    \label{equ:3}
    \chi^{2}=\sum_{i=1}^{N}\frac{(D_{i}-F_{i})^{2}}{F_{i}},
\end{equation}
where D$_i$ is the observed flux and F$_i$ is the flux predicted by the model, N is the number of bands.

From the derived column density values, we estimate the mass of the dusty clumps using the following expression:
\begin{equation}
    \label{equ:4}
    M_{clump}=\mu_{H_2}m_HArea_{pix}\sum N(H_2),
\end{equation}
where Area$_{pix}$ is the area of a pixel in cm$^2$. 

\begin{figure} [h!]
\centering
\includegraphics[width=1.0\columnwidth]{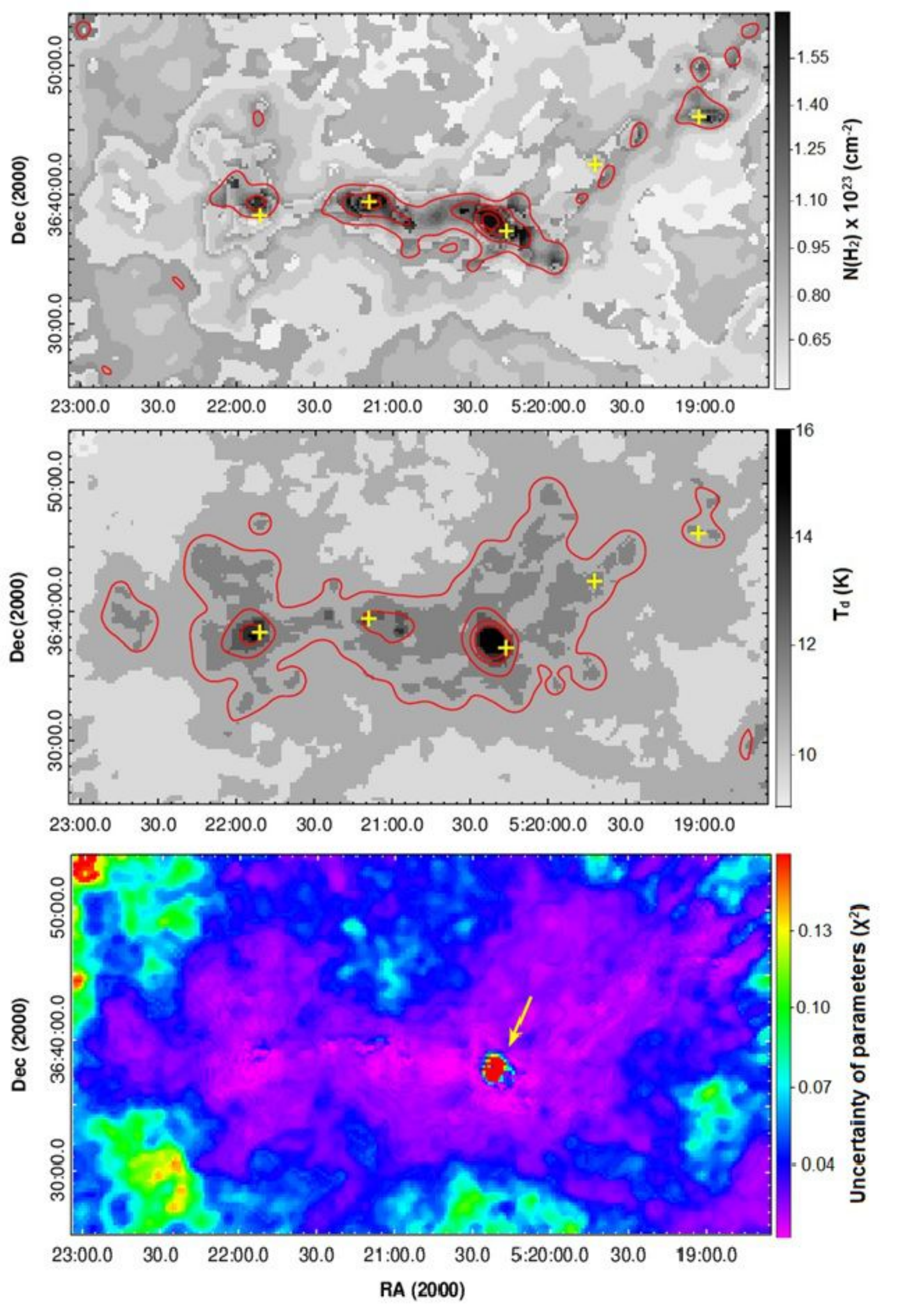}
\caption{Maps of N(H$_2$) column density (top panel), T$_d$ dust temperature (middle panel), and $\chi^2$ uncertainty of the parameters (bottom panel) of IRAS\,05168+3634 star-forming region. On the N(H$_2$) map the outer isodense corresponds to 0.9\,$\times$\,10$^{23}$\,cm$^{-2}$, and the interval between isodenses is $\sim\,$0.4\,$\times$\,10$^{23}$\,cm$^{-2}$. On the T$_d$ map, the outer isotherm corresponds to 11 K, and the interval between isotherms is 1 K. The arrow on the bottom panel indicates the positions of the region with a relatively high level of $\chi^2$. The positions of IRAS are marked by crosses.} 
\label{fig:2}
\end{figure}

Using the modified blackbody fitting on the \textit{Herschel} images obtained at 160, 250, and 500\,$\mu$m bands, we constructed maps of hydrogen column density and dust temperature in the considered star-forming region. The maps are presented in Figure \ref{fig:2}. On the N(H$_2$) and T$_d$ maps, the outer contours correspond to 0.9\,$\times$10$^{23}$\,cm$^{-2}$ (that exceeds the average N(H$_2$) of the surrounding molecular cloud by $\sigma$) and 11\,K, respectively. The maps show that the star-forming region, which includes five IRAS objects, clearly stands out against the background of the surrounding molecular cloud both with a higher density and temperature. The relatively hotter gas-dusty matter forms dense condensations around IRAS objects. An exception is the IRAS\,05162+3639 sub-region, near which on the N(H$_2$) map there is practically no region with a relatively higher density. In this regard, it should be noted that in A19, around IRAS\,05162+3639 source, in essence, no group of YSOs has been identified, but only 5 stars. In addition, in the vicinity of IRAS\,05162+3639 and IRAS\,05156+3643 sources, in contrast to the other three IRAS sources, no increase in temperature is observed. In general, in whole star-forming region T$_d$ varies from 11 to 24\,K, and N(H$_2$) - from $\sim$\,1.0 to 4.0\,$\times$\,10$^{23}$cm$^{-2}$. Detailed analysis of each sub-region is given below (see Section \ref{ss:3.2} and Table \ref{tbl:1}). 
 
\begin{figure} [h!]
\centering
\includegraphics[width=1.0\columnwidth]{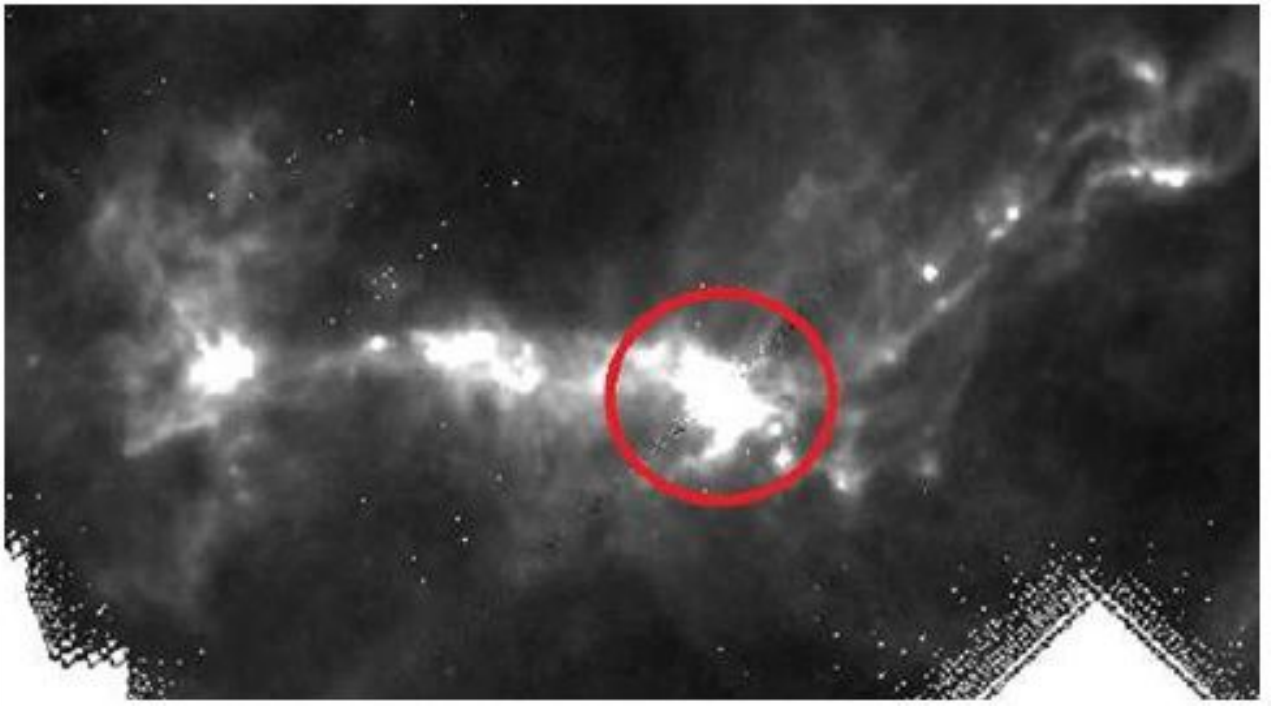}
\caption{\textit{Herschel} image at 250\,$\mu$m (2.5 Level). The red circle marks the location of bad pixels in the vicinity of IRAS\,05168+3634 source.} 
\label{fig:3}
\end{figure}

It should be noted that the parameters obtained with the modified blackbody model do not deviate from certain errors, which are related to a number of factors, including the quality of the images used and the uncertainty of calibration of \textit{Herschel} images. Besides, the errors are a strong function of the parameters’ values. To a large extent, the final result is also affected by the fact that star-forming regions show the structure of many scales, and so fitting a single column density and temperature to any point does not adequately represent the whole region \citep{Battersby2011}. Nevertheless, the $\chi^2$ map, presented in the bottom panel of Figure \ref{fig:2}, indicates that the fitting uncertainties within the star-forming region are relatively small. In general, the $\chi^2$ is about 0.01\,$-$\,0.02. However, directly in the center of IRAS\,05168+3634 sub-region, the $\chi^2$ value increases up to $\sim$\,1.5. This area is marked with an arrow in the bottom panel of  Figure \ref{fig:2}. In this case, the increase of $\chi^2$ is most likely explained by the quality of the images. Unfortunately, in all available \textit{Herschel} images, this star-forming region is located at the edge, where the image quality, especially in the 250\,$\mu$m channel, is not good enough. Figure \ref{fig:3} clearly shows that in the area of interest, in the image of Level 2.5, there are many bad pixels. Moreover, a whole strip of bad pixels passes through the bright condensation in IRAS\,05168+3634 sub-region, which whatever filtering cannot completely remove. This is most likely the reason for the increase in the $\chi^2$ value in this region.

 \begin{figure} [h!]
\centering
\includegraphics[width=1.0\columnwidth]{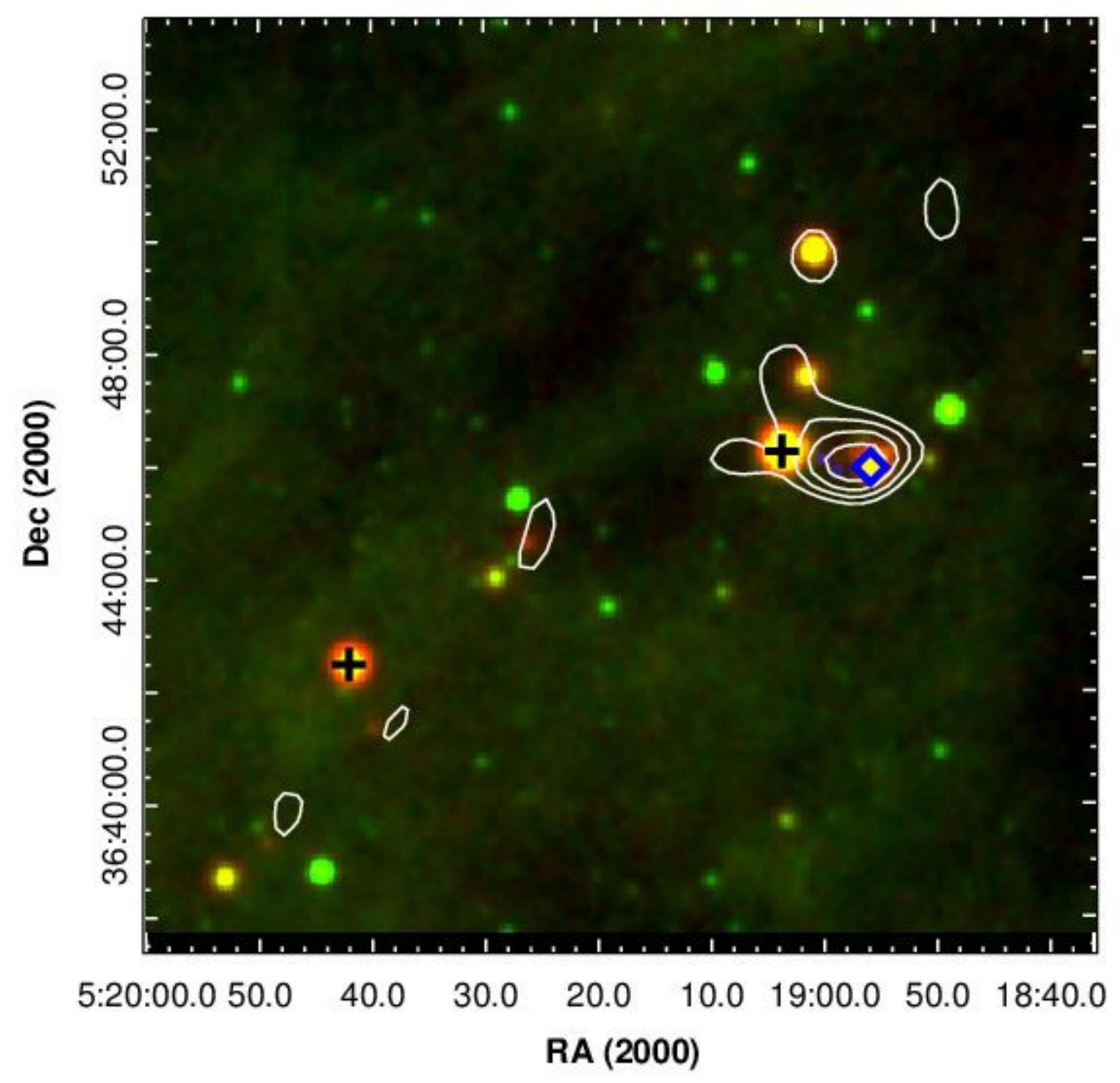}
 \caption{Colour-composite image of IRAS\,05156+3643 and IRAS\,05162+3639 sub-regions: W3 $-$ green, W4 $-$ red, white curves $-$ N(H$_2$) isodenses. In this and subsequent figures the outer isodense corresponds to 1.0\,$\times$\,10$^{23}$\,cm$^{-2}$ and interval between isodenses is 0.1\,$\times$\,10$^{23}$\,cm$^{-2}$. The positions of IRAS\,05156+3643 and 05162+3639 are marked by black crosses, BGPSv2\,G170.384-00.407 $-$ by blue diamond.} 
\label{fig:4}
 \end{figure}
 
\subsection{Individual Regions}
\label{ss:3.2}
\textit{\textbf{IRAS 05156+3643.}} Figure \ref{fig:4} shows the superposition of the N(H$_2$) enlarged contours of this sub-region with images W3 and W4. In these ranges, the key, brightest stellar members are clearly distinguishable. The maximal density and temperature in this sub-region are 1.6\,$\times$10$^{23}$\,cm$^{-2}$ and 12\,K, respectively. The masses of ISM within the isodence corresponding to 1.1\,$\times$10$^{23}$\,cm$^{-2}$, which exceeds the average surrounding medium hydrogen column density by 2$\sigma$, for this sub-region and all subsequent are given in Table \ref{tbl:1}. In the region with a maximum density, there is an object identified in 1.1\.mm Bolocam Galactic Plane Survey (BGPS) of dust emission. It is BGPSv2\,G170.384-00.407, marked by a blue diamond in Figure \ref{fig:4}. It should be noted that observations in the optically thin, minimally affected by temperature millimeter continuum provide one of the best methods to identify star-forming clumps \citep{Ginsburg2013}. It is quite expected that the BGPS object is localized precisely in the region of the N(H$_2$) maximum. According to A19, a young star with ID\,239, 4.4\,$\pm\,$0.3\,M$_{\odot}$ mass and evolutionary age of $\sim$\,10$^6$ years (here and after ID, mass, and evolutionary age are borrowed from Tables\,A.1 and A.3 in A19) is associated with it. The dust temperature reaches the maximum for this sub-region (12\,K) precisely at the location of BGPSv2\,G170.384-00.407 and IRAS\,05156+364 sources. IRAS\,05156+3643, also known as MSX G170.3964-00.3827, some shifted from the density maximum. The YSO with ID\,234 in the list in A19 is identified with it. It is the star with $\sim$\,3.8\,$\pm$\,3.6\,M$_{\odot}$ mass and evolutionary age of $\sim$\,10$^6$ years. According to coordinates and 2MASS JHK photometric data, this stellar object corresponds to the high proper-motion star LSPM\,1267-0099892 from the LSPM-NORTH catalog \citep{L2005}. Taking into account the high accuracy of parallax measurements and proper motion in \textit{Gaia}\,EDR3, we tried to identify the stellar object associated with IRAS\,05156+3643 in this database. According to the coordinates, it is the object with ID\,187051701258568960 in \textit{Gaia}\,EDR3, which parameters are presented in Table \ref{tbl:2} (see Section \ref{ss:4.2}). Unfortunately, the parallax measurement accuracy is not high enough and is $\varpi$/$\sigma_{\varpi}$\,=\,1.7, while for high quality parallaxes $\varpi$/$\sigma_{\varpi}$\,$>$\,5 \citep{Gaia2021}. However, according to the transformations of \citet{Bailer21}, which can infer meaningful distances even if the parallaxes are negative and/or the parallax accuracy is not enough, the low distance estimation is $\sim$\,2.2\,kpc, what is close to the parallax estimation for the star-forming region. Using the Galactocentric transformations of \citet{Johnson1987} (the similar transformation is also available in {\fontfamily{qcr}\selectfont
Astropy.coordinates}\footnote{\url{https://astropy-cjhang.readthedocs.io/en/latest/coordinates/index.html}} package), we compute peculiar velocity of this star for 2.2\,kpc distance: V$_{pec}$\,=\,57\,km/s. 

\begin{figure*} [h!]
\centering
\includegraphics[width=1.5\columnwidth]{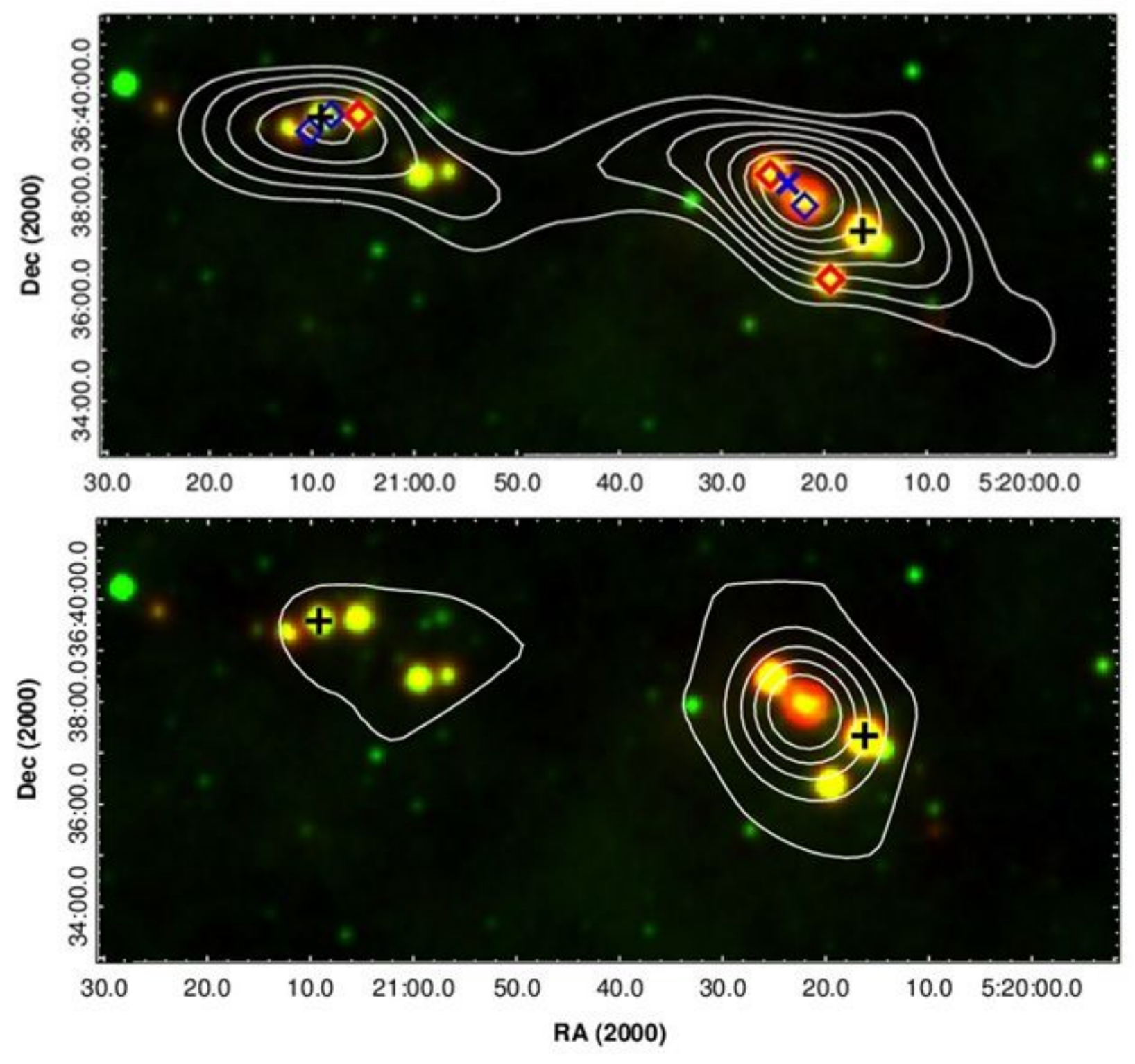}
 \caption{Colour-composite images of IRAS\,05168+3634 and IRAS\,05177+3636 sub-regions: W3 - green, W4 - red. \textit{Top panel}: white curves are N(H$_2$) isodenses. The outer isodense and interval between them are the same as in the previous figure. \textit{Bottom panel}: white curves are T$_d$ isotherms. The outer isotherm corresponds to 12\,K. The interval between isotherms is 1\,K. The positions of IRAS\,05156+3643 and 05162+3639 sources marked by black crosses, BGPS objects $-$ by diamonds, MSX objects $-$ by red diamonds, and pre-UC\,HII region $-$ by blue X.} 
\label{fig:5}
 \end{figure*}

The two most common scenarios to explain the existence of runaway stars and their high velocity are: (1) the supernova of the companion star in a massive binary \citep{Blaauw1961} and (2) the dynamical ejection from a young cluster in the early stages of evolution \citep{Poveda1967}. Apparently, the high peculiar velocity of the stellar object associated with IRAS\,05156+3643 is the result of (2) scenario. In this scenario, the runaway gains its velocity in the range 35\,$-$\,185\,km/s through a dynamical interaction with one or more other stars \citep{Poveda1967}.
\\  
\\ 
\textit{\textbf{IRAS\,05162+3639.}} In the vicinity of this IRAS source, we were unable to identify a region with a relatively high column density against the background of the surrounding molecular cloud. The dust temperature at the location of the IRAS source is 12\,K. Two stellar objects are associated with the IRAS source (A19). It is the object with ID\,190 (1.0\,$\pm$\,0.4\,M$_{\odot}$ mass and evolutionary age of $\sim$\,10$^5$ years), as well as the object with ID\,191  (0.6\,$\pm$\,0.1\,M$_{\odot}$ mass and evolutionary age of $\sim$\,10$^6$ years). Recall that A19 identified only five YSOs, which can be considered as potential members of this star-forming sub-region. In general, regarding their position on the colour-magnitude diagram and the results of the SED fitting tool in A19, there is no reason to assert that these stars are at an earlier evolutionary stage than the YSOs in other sub-regions, i.e. to suggest that the process of star formation is at an earlier stage here. Most likely, the initially not so significant concentration of ISM in this region contributed to the formation of only a few stellar objects.
\\  
\\ 
\textit{\textbf{IRAS\,05168+3634.}} The contours of the hydrogen column density and the dust temperature, superimposed on the W3 and W4 images, are shown in Figure \ref{fig:5}. The maximum values of density and temperature are 3.8\,$\times$\,10$^{23}$cm$^{-2}$ and 24\,K, respectively (see Table \ref{tbl:1}). It should be noted that in this region the values of these parameters are noticeably higher than in the others.

Figure \ref{fig:5} clearly shows that the localization of density and temperature maxima coincides. As in the previous region, the BGPSv2 object is located at the density and temperature maximum (blue diamond on Figure \ref{fig:5}). It is a dense core G170.661-00.249. According to the data in \citet{Zhang2005}, the CO outflow is centered near this dense core. Moreover, in \citet{Wolf-Chase2017} at least two collimated outflows of near-infrared MHOs have been detected, which follow the large-scale morphology of CO outflow. \citet{Wolf-Chase2017} suggest that their source may be one YSO, identified in the WISE database as J052022.03+363756.5, which is only 6\,arcsec away from the BGPS object. It should be noted that the direction of the isodences (PA is about 35$^{\circ}$) is almost identical to that of CO outflow (PA is about 40$^{\circ}$). According to the data in A19, the closest to G170.661-00.249 and J052022.03+363756.5 (on a distance 6\,$\arcsec$ $-$ 14\,$\arcsec$) two objects are located: ID\,156 and ID\,175. Unfortunately, their parameters were not determined by the SED fitting tool. According to photometric data, these stars are objects of the II and I evolutionary classes, respectively. It can be assumed that each of them can be the source of one of two near-infrared MHOs outflows.

According to previous studies \citep{Molinari1996,Wang2009}, northeast of BGPSv2\,G170.661-00.249, at a distance of about 30\,$\arcsec$ it was detected a high-mass star-forming region in pre-UC\,HII phase, which appears to be associated with the infrared source (blue X on Figure \ref{fig:5}). The radius of the pre-UC\,HII region is more than 8\,$\arcsec$ \citep{Wang2009}. In A19, the stellar object with ID\,186 is the closest to the pre-UC\,HII region ($\sim$\,11\,$\arcsec$). Unfortunately, its parameters (mass and evolutionary age) have not been determined. However, according to photometric data, this YSO has a very significant infrared excess (J\,-\,K\,=\,4.7), which may indicate its very early evolutionary stage. Although the distance is slightly larger than the radius, this object can presumably be considered as an infrared source of the pre-UC\,HII region. Of course, it is possible that the central protostellar object of pre-UC\,HII formation is at a very early evolutionary stage and is so embedded that it has not been identified in A19 in NIR and MIR bands. The closest MSX object G170.6589-00.2334, which is associated with intermediate-mass YSO (ID\,177, 1.9\,$\pm$\,0.7\,M$_{\odot}$) with an evolutionary age of $\sim$\,10$^5$ years, is located much further, at a distance of about 24\,$\arcsec$ to the northeast (red diamond on Figure \ref{fig:5}).

IRAS\,05168+3634 source (MSX\,G170.6575-00.2685), with which the intermediate-mass YSO (5.1\,$\pm$\,0.3\,M$_{\odot}$) with an evolutionary age of about 10$^6$ years is associated (ID\,183 in A19), is displaced from the position of the N(H$_2$) and T$_d$ maxima (black cross on Figure \ref{fig:5}). Directly in its vicinity, the value of N(H$_2$) is 1.4\,$\times$\,10$^{23}$\,cm$^{-2}$, which exceeds the density estimate from $^{13}$CO data J\,=\,2\,$-$\,1 observations \citep[2.1\,$\times\,10^{22}$\,cm$^{-2}$,][]{Wang2009}. Another MSX object, G170.6758-00.2691, marked by a red diamond, is located almost on the periphery of the cluster, southeast of IRAS\,05168+3634. It is also associated with an intermediate-mass YSO (4.9\,$\pm$\,0.5\,M$_{\odot}$) with an evolutionary age of $\sim$\,10$^6$ years (ID\,182 in A19).

IRAS\,05168+3634 sub-region is entirely embedded in the "non-sourceless" (\textit{there is at least one MSX source within
on the beam size from the core centre}) $^{13}$CO core \citep{Guan2008}. According to the results presented in \citet{Guan2008}, two more $^{13}$CO cores, but already "sourceless" (\textit{the core is sourceless by MSX and IRAS catalogs}), are located between IRAS\,05168+3634 and IRAS\,05177+3636 sub-regions, where no stellar objects belonging to the star-forming region were detected (A19). Another "non-sourceless" $^{13}$CO core was identified in the IRAS\,05177+3636 sub-region, which will be discussed below.
\\

\textit{\textbf{IRAS\,05177+3636.}} The contours of the hydrogen column density and dust temperature superimposed on W3 and W4 images of this star-forming region are also presented in Figure \ref{fig:5}. The maximum density here is 2.3\,$\times$\,10$^{23}$\,cm$^{-2}$. The isotherm corresponds to T$_d$\,=\,12\,K. Inside it, the T$_d$ reaches its maximum (13\,K) in the vicinity of the brightest stellar objects. The IRAS object (also known as MSX\,G170.7268-00.1012) is associated with the intermediate-mass young star with 2.6\,$\pm$\,1.3\,M$_{\odot}$ mass and an evolutionary age of less than 10$^6$ years (ID\,97 in A19). The second MSX object (red diamond in Figure \ref{fig:5}), G170.7196-00.1118, is also associated with intermediate-mass YSO (4.5\,$\pm$\,0.7\,M$_{\odot}$) and an evolutionary age also less than 10$^6$ years (ID\,128 in A19).

In contrast to the previous case, in this star-forming region, the IRAS source is located practically in the center of the cluster, in the zone of maximum density. There are two BGPSv2 objects in its immediate vicinity (blue diamonds in Figure \ref{fig:5}): G170.724-00.105 and G170.733-00.103. They are associated with three objects of the submillimeter continuum detected by SCUBA (450\,$\mu$m, 850\,$\mu$m): J052109.2+363934, J052106.7+363946, and J052112.2+ 363916 \citep{Di2008}. Besides the IRAS object, two more YSOs are located closest to the BGPSv2 objects: the intermediate-mass star (3.7\,$\pm$\,0.7\,M$_{\odot}$) with an evolutionary age of less than 10$^6$ years (ID\,85 in A19) and the low-mass star (0.3\,$\pm$\,0.2\,M$_{\odot}$), apparently at a very early evolutionary stage (ID\,96 in A19). It has an evolutionary age of less than 10$^5$ years.

\begin{figure*} [h!]
\centering
\includegraphics[width=1.9\columnwidth]{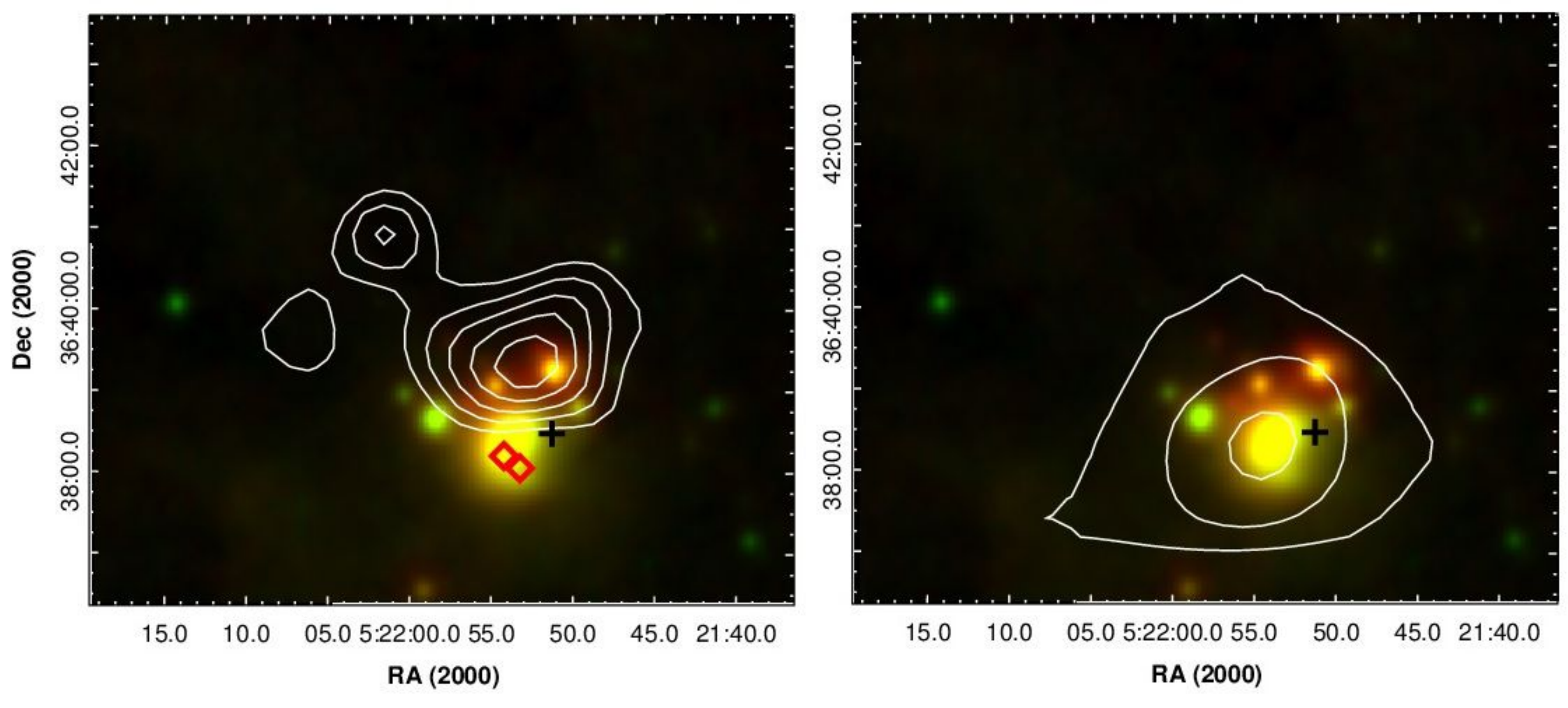}
 \caption{Colour-composite images of IRAS\,05184+3635 sub-region: W3 - green, W4 - red. \textit{Left panel}: white curves are N(H$_2$) isodenses. \textit{Right panel}: white curves are T$_d$ isotherms. The outer isodense and isotherm, as well as the intervals between them, are the same as in the previous figure. The position of IRAS\,05184+3635 source is marked by a black cross, of the MSX objects - by red diamonds.} 
\label{fig:6}
 \end{figure*}

We would also like to note that the search for molecular outflows based on $^{12}$CO observations in \citet{Casoli1986} did not yield a positive result.
\\
\\
\textit{\textbf{IRAS\,05184+3635.}} \citet{Lundquist2014}, on the basis of extended emission at 12\,$\mu$m (W3) and 22\,$\mu$m (W4), characterized this star-forming region as a region with blobs/shells or circular/elliptical morphology. This conclusion is consistent with the almost spherical isodenses and isoterms of the star-forming region, which are shown in Figure \ref{fig:6}. The maximum values of density and temperature are 1.5\,$\times$\,10$^{23}$\,cm$^{-2}$ and 15\,K. Unlike other sub-regions, there is a significant shift between the density and temperature maxima. Moreover, all key, bright stellar objects, namely the IRAS (black cross) and two MSX sources (red diamonds), are localized mainly in the region of the temperature maximum. A young star of intermediate-mass (3.1\,$\pm$\,0.6\,M$_{\odot}$) and evolutionary age of $\sim$\,10$^6$ years (ID\,31 in A19) is associated with the IRAS source (or MSX\,G170.8276+00). There are two more MSX sources in its vicinity: G170.8321+00.0045 and G170.8319+00.0086. According to the lists in A19, in the vicinity of these sources, in addition to the ID\,31 stellar object, within a radius of 11\,$\arcsec$ $-$ 14\,$\arcsec$, only one stellar object of the II evolutionary class has been identified (ID\,11 in A19), which mass and evolutionary age have not been determined. We would also like to note that no BGPS objects have been identified here. In addition, no stellar objects were identified in A19 in the vicinity of the second, northeastern maximum of density.

\begin{figure*} [h!]
\centering
\includegraphics[width=1.9\columnwidth]{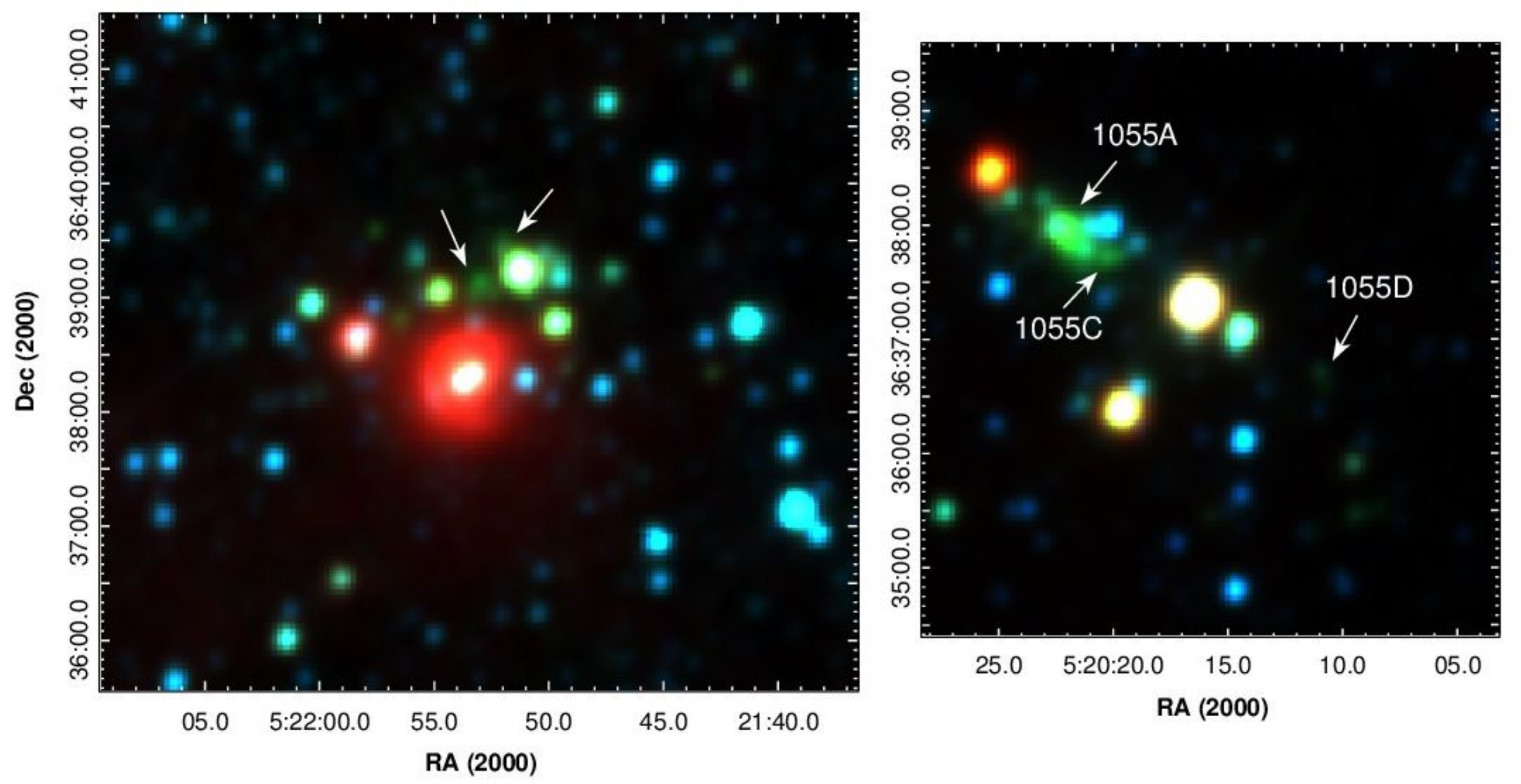}
 \caption{Colour-composite images of IRAS\,05184+3635 (left panel) and IRAS\,05168+3634 (right panel) sub-regions: W1 (blue), W2 (green), and W3 (red). The mid-infrared tracers of outflow in IRAS\,05184+3635 sub-region are marked by arrows. The near-infrared MHOs in the IRAS\,05168+3634 sub-region are labeled.} 
\label{fig:7}
 \end{figure*}

There is no information in previous studies on the presence of collimated outflow in this sub-region. In addition, the search for water maser emission did not return a positive result \citep{Wouterloot1993}. Nevertheless, by comparing W1, W2, and W3 images in the vicinity of IRAS\,05184+3635 source, we were able to identify a diffuse structure with excess emissions in the W2 band (see Figure \ref{fig:7}, left panel). In the diffuse structure, two concentrations are clearly distinguishable, which are indicated by arrows. It is known that stellar outflows have strong molecular hydrogen emission, which is detectable in the visible, NIR, and MIR. In MIR shocked H$_2$ emission appears particularly strong in the 4.5\,$\mu$m Spitzer Infrared Array Camera (IRAC) band \citep[e. g.][]{Noriega2004,Ybarra2009,Teixeira2008}. Unfortunately, for this star-forming region, IRAC images were obtained in only two bands: 3.6 µm and 4.5 µm, and there are no longer wavelength images for comparison. Therefore, we used the colour-composite WISE images, obtained in 3.4, 4.6, and 12.0\,$\mu$m bands. Of course, the resolution of WISE images is inferior to IRAC data. Nevertheless, the excellent sensitivity of the WISE data enables the detection of many faint features. Undoubtedly, a shocked nature of the obtained emission objects requires a spectroscopic follow-up.

For comparison, we also present the W1, W2, and W3 colour-composite image for IRAS\,05168+3634 sub-region (see Figure \ref{fig:7}, right panel). As noted above in \citet{Wolf-Chase2017}, there were presented the results of a narrow-band NIR imaging survey for MHOs toward IRAS\,05168+3634 sub-region. It was revealed that there is collimated chains of MHOs. We detect that several MHOs, marked in Figure \ref{fig:7}, also have excess emission in W2 band. This fact increases the confidence that the concentrations of excess emissions in the W2 band in IRAS\,05184+3635 sub-region are indeed traces of outflow. Their source is presumably a stellar object 31 in A19. Moreover, we want to draw attention to the fact that, as in IRAS\,05168+3634, the direction of the alleged outflow coincides with that of the external isodenses.
 
It should also be noted that the search for MIR tracers of outflow carried out in other sub-regions did not yield a positive result.

\section{Discussion}\label{s:4}
\subsection{ISM and stellar content} \label{ss:4.1}
Table \ref{tbl:1} shows the main parameters of ISM in the IRAS sub-regions, including N(H$_2$), T$_d$, and ISM masses. For comparison, we also included in Table  \ref{tbl:1} some parameters of the stellar content (sub-region's radius, YSOs number, surface stellar density, range of stellar masses) borrowed from A19. Taking into account the number of identified stellar objects, the average value of their masses, as well as the ISM mass in the surrounding cloud, we find that star formation efficiency (SFE\,=\,M$_{stars}$/(M$_{stars}$\,+\,M$_{ISM}$)) in all sub-regions is less than 0.1\%. Even if we take into account that not all stellar objects were detected, the SFE value still indicates an active star formation process in the entire region.

\begin{table*} [h!]
\small
\caption{Parameters of the stellar content and ISM in IRAS sub-regions} 
\resizebox{1\textwidth}{!}{
\label{tbl:1}
\begin{tabular}{cccccc}
 \tableline 
 \textbf{Parameter / IRAS} &	\textbf{05156+3643} &	\textbf{05162+3639} &	\textbf{05168+3634} &	\textbf{05177+3636} & \textbf{05184+3635} \\  \hline
 Radius (arcmin)	& 2.8 & 	0.25 &	3.0	& 3.5 &	2.5 \\
 YSO's number	& 47 &	5 &	57	& 79 &	52 \\
 n$_{star}$\,(arcmin$^{-2}$) &	1.9 &	25.5 &	2.0	& 2.1 &	2.6 \\
 Class I (\%)	& 20 &	- &	43	& 28 &	21 \\
 Stellar masses (M$_{\odot}$)	& 0.2 $-$ 1.6 &	- &	0.5 $-$ 2.5	& 0.2 $-$ 2.2 &	0.3 $-$ 1.5\\
  N(H$_2$) ($\times$10$^{23}$\,cm$^{-2}$) &	1.1 $-$ 1.6	& 0.9 $-$ 1.0 &	1.1 $-$ 3.8	& 1.1 $-$ 2.3 &	1.1 $-$ 1.5 \\
  T$_d$ (K)	& 11 $-$ 12	& 11 $-$ 12	& 11 $-$ 24	& 12 $-$ 13	& 12 $-$ 15 \\
  $^*$ISM Mass (M$_{\odot}$) &	(1.7\,$\pm$\,0.05)\,$\times$\,10$^4$ &	- &	(2.1\,$\pm$\,1.6)\,$\times$\,10$^5$ &	(9.2\,$\pm$\,0.3)\,$\times$\,10$^4$ & (4.0\,$\pm$\,0.1)\,$\times$\,10$^{4}$ \\
 \tableline 
\end{tabular}
}

\tablenotetext{}{$^*$ISM masses are determined in the area limited by the isodence corresponding to 1.1\,$\times$\,10$^{23}$\,cm$^{-2}$ value, which exceeds the average hydrogen column density in surrounding medium by 2$\sigma$. The error bars are due to the average value of the uncertainty of parameters ($\chi^2$) in the sub-regions and the uncertainty in estimating the distance to the star-forming region.}

\end{table*}

The BGPSv2 objects identified in the three sub-regions (IRAS\,05156+3643, 05168+3634, and 05177+3636) are located at maximum density and temperature. On the one hand, this is an expected fact, since these 1.1\,mm sources are good indicators of dust emission and, therefore, star-forming clumps. On the other hand, this confirms the reliability of our results. However, another key, brightest stellar objects, including associated with the IRAS and MSX sources, localized as regions of N(H$_2$) and T$_d$ maxima, as well as significantly displaced from them. We also want to note that in most cases, that the evolutionary age of associated with the BGPSv2, IRAS and MSX sources intermediate-mass stellar objects is about 10$^6$ years.

The dust temperature, excluding IRAS\,05168+3634 sub-region, is only a few degrees higher than that in the surrounding molecular cloud. According to the results in A19, no high-mass stars were found in the cluster, which is probably the reason for the relatively low temperature in this region. A small increase in T$_d$ to 15\,K in the IRAS\,05184+3635 sub-region, most likely, can be explained by the highest stellar density. Since only 5 stellar objects were detected in IRAS\,05162+3639 sub-region, we do not take it into account in the comparison. As mentioned above, in IRAS\,05168+3634 sub-region, it was identified a high-mass star-forming region in pre-UC\,HII phase \citep{Molinari1996}, which could be the reason why the dust temperature in this region is significantly, almost 10\,K, higher.

Regarding the distribution of N(H$_2$), we would like to draw attention to one more fact. In the two sub-regions, IRAS\,05168+3634 and 05184+3635, where the outflow was detected, its orientation is well correlated with the isodenses' directions. This is in good agreement with the findings in \citet{Arce2010,Mottram2012} that outflows from newly formed stars inject momentum and energy into the surrounding molecular cloud, which may affect the morphology of the ISM, including the distribution of the hydrogen column density in the local environment.

\subsection{Distance of the star-forming region}
\label{ss:4.2}
\begin{figure} [tb]
\centering
\includegraphics[width=1\columnwidth]{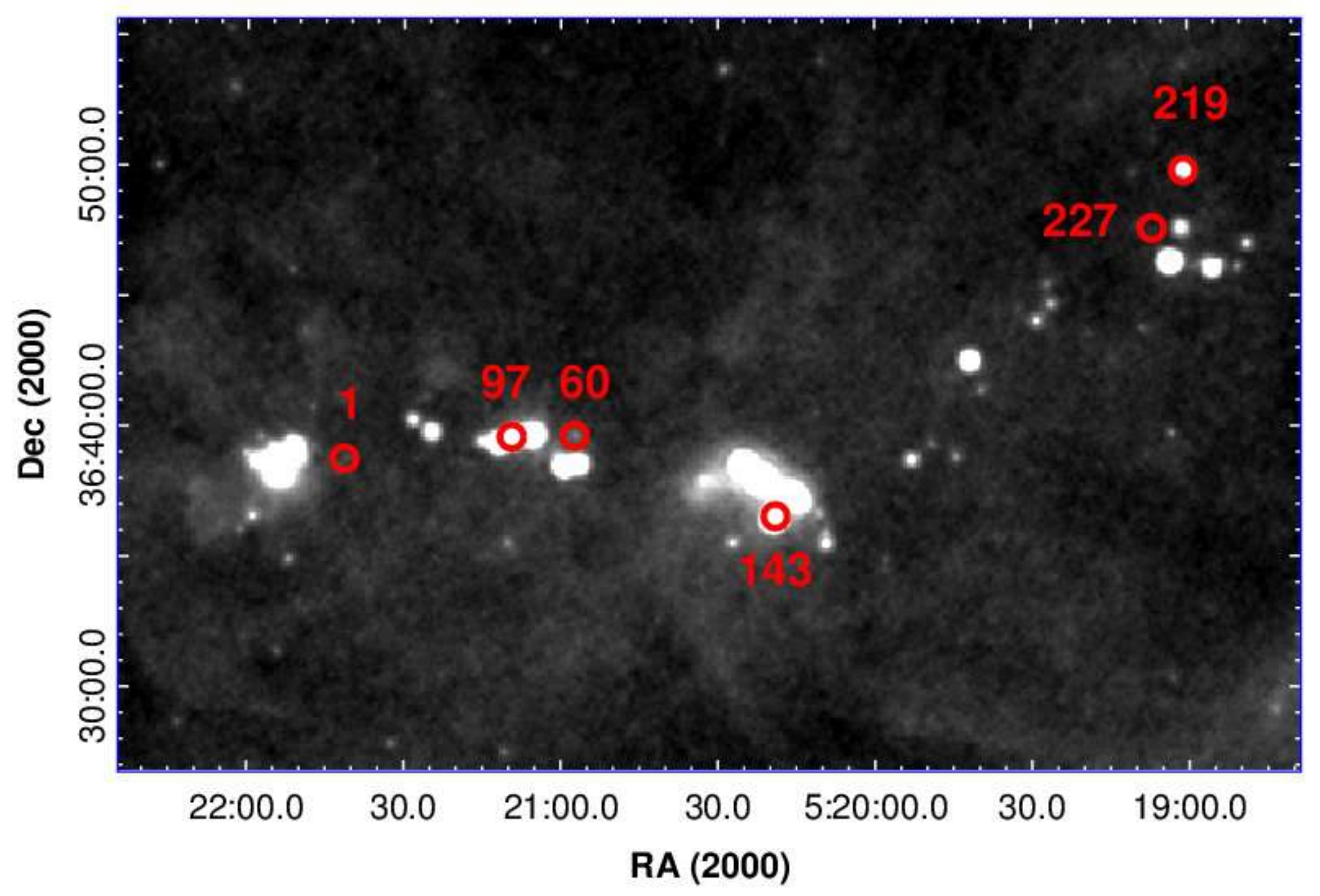}
 \caption{W4 image of the star-forming region. The positions (red circles) and ID numbers of 6 stars, which, according to the \textit{Gaia}\,EDR3 data, are located at a distance of $\sim$\,2\,kpc (see Table\,\ref{tbl:2}) are marked.}
\label{fig:9}
 \end{figure}

As noted above, in Section \ref{ss:3.1}, IRAS\,05156+3643 or the 234 star in A19 was identified in the \textit{Gaia}\,EDR3 database. The parallax measurement accuracy was not high enough ($\varpi$/$\sigma_{\varpi}$\,=\,1.7), however, according to the transformations of \citet{Bailer21} the low distance estimation is $\sim$\,2.2\,kpc, what is close to the parallax estimation with VERA \citep[$\sim$\,1.9\,kpc,][]{Sakai2012}. This led to an attempt to identify the remaining stars from the list in A19 in the \textit{Gaia}\,EDR3 database. In total, we were able to identify 65 objects, but only for 11 of them the parallax measurement accuracy is high enough ($\varpi$/$\sigma_{\varpi}$\,$>$\,5). The data of these 11 objects are shown in Table\,\ref{tbl:2}. Such a small number of objects identified with sufficient accuracy is quite expected. The embedded in a dense ISM cluster contains YSOs, which are surrounded by optically thick circumstellar disks and envelopes. In combination, this makes cluster members difficult to access optically. The number of identified objects is not sufficient for a detailed statistical study, but nevertheless allows us to draw some conclusions about the distance of the star-forming region as a whole. Distances of 6 out of 11 stars in Table\,\ref{tbl:2} are close to the parallax estimation for IRAS\,05168+3634. Their positions and ID numbers in Table\,\ref{tbl:2} are shown on Figure\,\ref{fig:9}. We can see that these stellar objects are scattered in almost all sub-regions, with the exception of only IRAS\,05162+3639. Moreover, one of them, the 97$^{th}$, is associated with IRAS\,05177+3636. We would like to remind that direct measurements of distances were made only for IRAS\,05168+3634. The conclusion that all sub-regions belong to a single star-forming region in A19 was made mainly on the statistical analysis of the distribution of YSOs of the I evolutionary class. The result obtained from \textit{Gaia}\,EDR3 data can be considered as one more argument in favor of the fact that all sub-regions are embedded in the single molecular cloud and belong to the same star-forming region, which is located at a distance of $\sim$\,1.9\,kpc.

Unlike IRAS\,05156+3643, the V$_{pec}$ values for these 11 stars (see Table \ref{tbl:2}) do not give grounds for considering them as runaway objects. Although it should be noted that the peculiar velocities of stellar objects are incomplete, since their radial velocities are not available.

\begin{table*} [tb]
\small
\centering
\caption{\textit{Gaia} EDR3 data} 
\resizebox{1\textwidth}{!}{
\label{tbl:2}
\begin{tabular}{cccccccc}
\tableline  
\textbf{N} & \textbf{\textit{Gaia}\,EDR3} & \textbf{$\varpi$\,(mas)}  & \textbf{$\sigma_{\varpi}$\,(mas)} &	\textbf{PM\,(mas/yr)} & \textbf{LD\,(pc)} & \textbf{HD\,(pc)} &	\textbf{V$_{pec}$\,(km/sec)}
\\
\hline
(1) & (2) & (3) & (4) & (5) & (6) & (7) & (8) \\
\hline
1	& 184011624623256704 &	0.520 &	0.025 &	3.054 &	1780 &	1963 &	26\,$\pm$\,1.5 \\
14	& 184013166513879808	& 7.795	& 1.234	& 18.875	& 116	& 162	& 16\,$\pm$\,1.0 \\
60	& 184053612223579136	& 0.558	& 0.052	& 2.846	& 1597	& 1936 &	20\,$\pm$\,2.0 \\
82	& 184059620881914880 &	2.443	& 0.390	& 1.875	& 376 &	516	& 10\,$\pm$\,0.5  \\
91	& 184053436128920448	& 1.683	& 0.261	& 2.755	& 532	& 758 &	11\,$\pm$\,0.3 \\
97	& 184059350299876608 &	0.541	& 0.087	& 2.525	& 1608	& 2290 &	17\,$\pm$\,4.0 \\
101	& 184053401769173888	& 0.965	& 0.130	& 2.864 &	897	& 1139 &	12\,$\pm$\,0.5 \\
132	& 184055536368961792	& 1.279	& 0.019	& 11.533 & 	744 &	769 &	36\,$\pm$\,1.0 \\
143 &	184054746094985984 &	0.505	& 0.053	& 2.819	& 1749 & 	2170	& 22\,$\pm$\,3.0 \\
219	& 187052182295036672	& 0.648	& 0.092	& 2.854	& 1319	& 1820 &	17\,$\pm$\,3.0 \\
227	& 187051877354793984 &	0.583	& 0.109	& 3.428	& 1534	& 2662 &	22 $-$ 40\\
\hline
234	& 187051701258568960	& 0.401	& 0.236	& 3.428	& 2224 &	4780 &	51$-$110 \\

\tableline 
\end{tabular}
}
\tablenotetext{}{(1) $-$ a serial number of stars in A(19), (2) $-$ a  source designation in \textit{Gaia}\,EDR3, (3) - (5) $-$ parallax, parallax error and proper motion obtained from \textit{Gaia}\,EDR3 database, (6), (7) $-$ low and high distances obtained through \citet{Bailer21}, (8) $-$ peculiar velocity obtained by using the Galactocentric transformations of \citet{Johnson1987}.}

\end{table*}

\subsection{Origin of the star-forming region} \label{ss:4.3}
\begin{figure} [tb]
\centering
\includegraphics[width=1\columnwidth]{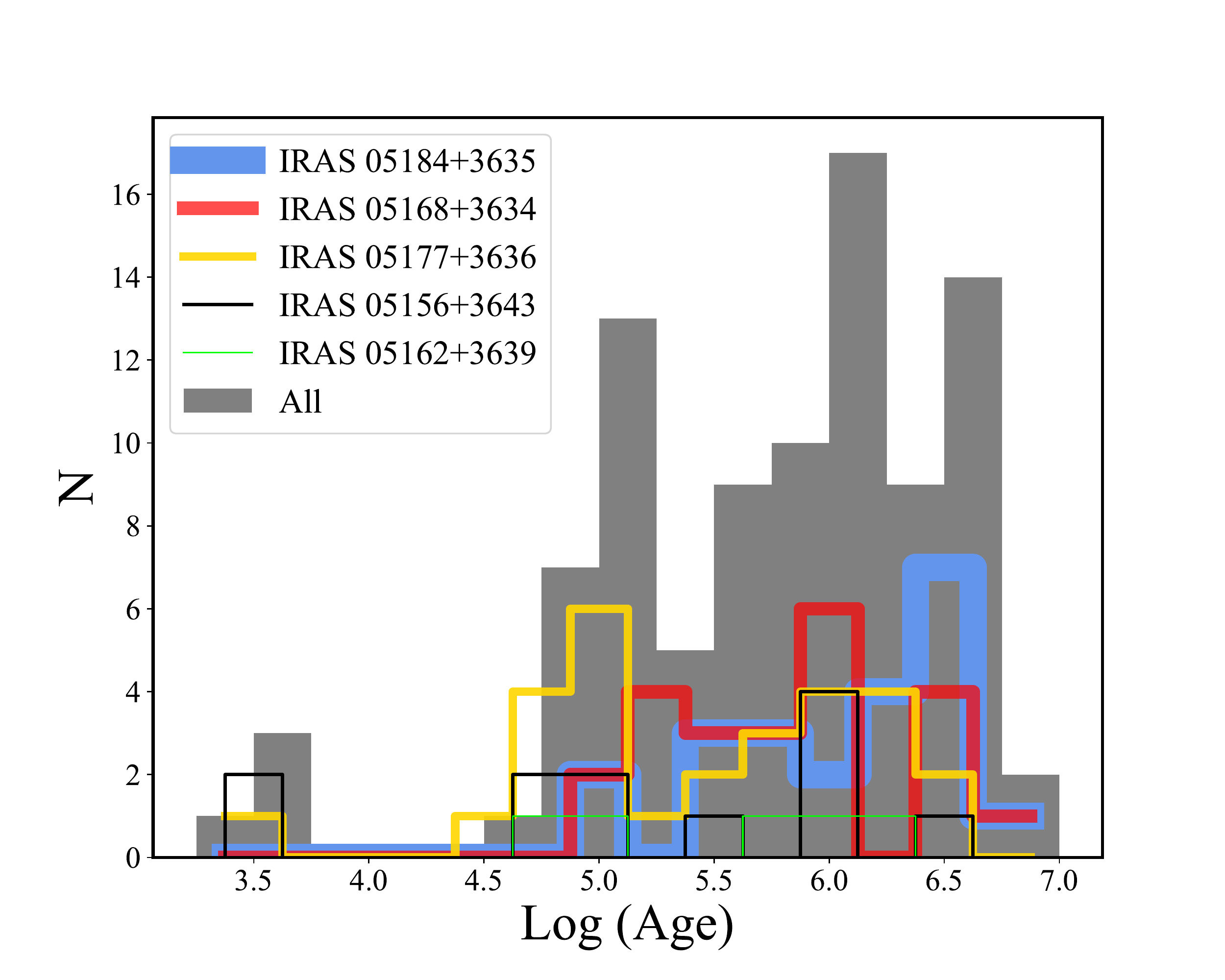}
 \caption{Histogram of the evolutionary ages determined by the SED fitting tool for members of the IRAS sub-regions. The bin size corresponds to Log (Age)\,=\,0.25.} 
\label{fig:8}
 \end{figure}

If we classify the evolutionary stage of sub-regions in relation to the percentage of Class\,I and Class\,II YSOs, then IRAS\,05168+3634 sub-region can be considered the “youngest". The next “youngest" sub-region by the percentage of Class\,I stellar objects is IRAS\,05177+3636. At the same time, IRAS\,05168+3634 and IRAS\,05177+3636 sub-regions have the highest hydrogen column density and ISM mass. The sub-regions around IRAS\,05184+3635 and IRAS\,05156+3643 in the outer part of the molecular cloud, according to the same criterion, are the “oldest". It should be noted once again that since there is no real concentration around IRAS\,05162+3639, we do not consider this region. On the other hand, if we consider the position of stellar objects on the colour-magnitude diagram in A19, it is clearly seen that stars from all sub-regions have a wide spread relative to the colour index (J-K)$_0$ and, therefore, isochrones. For clarity, we also built a histogram of the evolutionary ages of sub-regions' members, defined using the SED fitting tool and borrowed from A19. Unfortunately, this information is available only for 50\% of all identified members of the cluster. A histogram for a distance of 1.9\,kpc is shown in Figure \ref{fig:8}. The histogram also clearly shows that the evolutionary age of stars in all sub-regions has very wide spread. The overwhelming majority of them are located in the range from 10$^5$ to 10$^7$ years.

On the basis of the above, it can be assumed that, in general, the star formation process in the considered region is sequential. Moreover, in those sub-regions where the mass of the initial, parent molecular cloud is greater, this process is more active and presumably has a longer duration. As a result, at the moment, the percentage of younger stars could be more. Based on the results obtained from different regions of star formation, it seems that some clusters owe their origin to an external triggering shock and other clusters can be formed as independent condensations in molecular clouds \citep[e.g.][]{Elmegreen,Zinnecker1993}. If the star formation is triggered, the age spread of new generation stars should be small, while in self-initiated condensations the age spread of young stellar clusters is large \citep{Zinnecker1993}. Indeed, many careful studies of star-forming regions find no clear evidence or at most very moderate age spreads of several Myr \citep{Lee2007}. Moreover, in some cases, the measured age spread is less than the crossing time, which is well consistent with the scenario of the fast trigger nature of the star formation process \citep[][and ref. therein]{Preibisch2012}. However, the age spread of the considered star-forming region is much larger, and, therefore, it can be concluded that the stellar population is formed as a result of independent condensations in the parent molecular cloud.

\section{Conclusion} \label{s:5}
The main research tasks in the star-forming region, which includes IRAS 05156+3643, 05162+3639, 05168+3634, 05177+3636, and 05184+3635 sources, are the determination of ISM physical parameters (N(H$_2$) hydrogen column density and T$_d$ dust temperature), as well as their distribution. We also provide a comparative analysis of the properties of the ISM and YSOs. To determine the distribution of N(H$_2$) and T$_d$, we used the Modified blackbody fitting on \textit{Herschel} images obtained in four bands: 160, 250, 350, and 500\,$\mu$m. An analysis of the results revealed the following: 

\begin{itemize}
    \item The gas-dust matter has an inhomogeneous structure, forming relativity dense condensations around the majority of IRAS sources, which are interconnected by a filament structure.
     \item In general, in sub-regions T$_d$ varies from 11 to 24\,K, and N(H$_2$) - from 1.0 to 4.0\,$\times$\,10$^{23}$\,cm$^{-2}$. The masses of the ISM vary from 1.7\,$\times$\,10$^4$ to 2.1\,$\times$\,10$^5$ M$_{\odot}$. 
      \item The SFE in the sub-regions is less than 0.1\%.
      \item All identified in this star-forming region BGPSv2 objects are located at the density maximum. 
      \item In two sub-regions, IRAS\,05168+3634 and  05184+3635, where the outflows were detected, their orientations correlate well with isodenses directions.
      \item The intermediate-mass YSO associated with IRAS\,05156 +3643 source is a runaway star with V$_{pec}$ equal to $\sim$\,50\,km/s.
      \item On the \textit{Gaia}\,EDR3 database, it can be assumed that all sub-regions are embedded in the single molecular cloud and belong to the same star-forming region, which is located at a distance of $\sim$\,1.9\,kpc.
      \item The sub-regions with the highest hydrogen column density and ISM mass have the largest percentage of YSOs with Class\,I evolutionary stage.
\end{itemize}

Based on the fact that the evolutionary ages of stars in all sub-regions have a very wide spread  (from 10$^5$ to 10$^7$ years), we assumed that in the considered region the stellar population is formed as a result of independent condensations in the parent molecular cloud, and, therefore, the star formation process is sequential. In sub-regions where the mass of the initial, parent molecular cloud is greater, this process is most likely more active.

%
\acknowledgments 
We are very grateful to the anonymous referee for the helpful comments and suggestions. This work was made possible by a research grant number №\,21AG-1C044 from Science Committee of Ministry of Education, Science, Culture and Sports RA. This work also was made possible in part by a research grant astroex-5525 from the Yervant Terzian Armenian National Science and Education Fund (ANSEF) based in New York, USA. This work is based on observations made with Herschel, a European Space Agency Cornerstone Mission with significant participation by NASA. Support for this work was provided by NASA through an award issued by JPL/Caltech.This publication also makes use of data products from the Wide-field Infrared Survey Explorer, which is a joint project of the University of California, Los Angeles, and the Jet Propulsion Laboratory/California Institute of Technology, funded by the National Aeronautics and Space Administration.We gratefully acknowledge the use of data from the NASA/IPAC Infrared Science Archive, which is operated by the Jet Propulsion Laboratory, California Institute of Technology, under contract with the National Aeronautics and Space Administration.This work also presents results from the European Space Agency (ESA) space mission Gaia. Gaia data are being processed by the Gaia Data Processing and Analysis Consortium (DPAC). Funding for the DPAC is provided by national institutions, in particular the institutions participating in the Gaia MultiLateral Agreement (MLA).

%

%
%

%

%


%
\bibliographystyle{spr-mp-nameyear-cnd}  
\bibliography{biblio-u1}                

%

\end{document}